\documentclass{v19cosmo}
\usepackage{amssymb}
\usepackage{amsmath}
\usepackage{cite}
\usepackage{graphicx}
\def\be{\begin{equation}}
\def\ee{\end{equation}}
\def\bea{\begin{eqnarray}}
\def\eea{\end{eqnarray}}

\newcommand{\Photo}{\includegraphics[height=35mm]{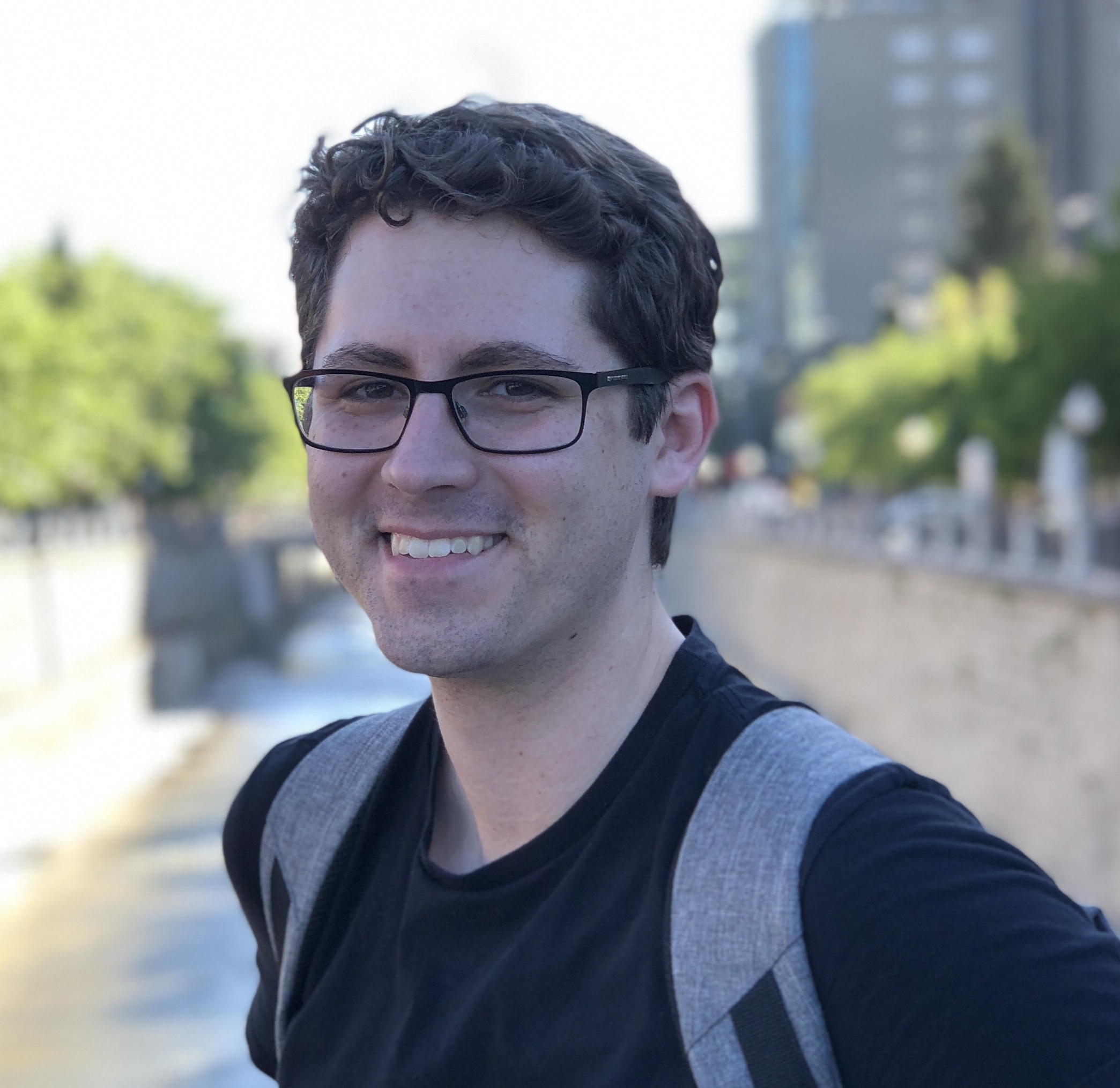}}

\begin{document}
\vspace*{4cm}
\title{Constraints on Dark Energy from Inflation and the Swampland Conjectures\footnote{Contribution to proceedings for Rencontres du Vietnam, August 2019.}}

\author{ Jacob M. Leedom }

\address{Department of Physics, University of California, Berkeley, California 94720, USA \\ Theoretical Physics Group, Lawrence Berkeley National Laboratory, Berkeley, California, 94720, USA}

\maketitle\abstracts{We discuss the prospects of measuring deviations of the dark energy equation of state from $w=-1$ by using the swampland conjectures to relate inflationary models to quintessence scenarios. This note is based on work done by the author with H. Murayama and C. Chiang~\cite{Chiang:2018lqx}.}

\section{The Landscape and the Swampland}

	The consistent coupling of gravity to quantum field theories remains one of the great open questions in theoretical physics.
	Recent progress has established that the space of effective field theories (EFTs) coupled to gravity is divided into two subsets - the landscape and the swampland~\cite{Vafa:2005ui,Ooguri:2006in}. The landscape consists of all EFTs that have a consistent UV quantum gravity completion. Such EFTs are typically considered as arising from various compactifications of string theory. On the other hand, the swampland is the set of all EFTs coupled to gravity that appear viable but do not have a consistent UV completion. 

\begin{figure}[tbh]
\centering
\includegraphics[width=0.3\textwidth]{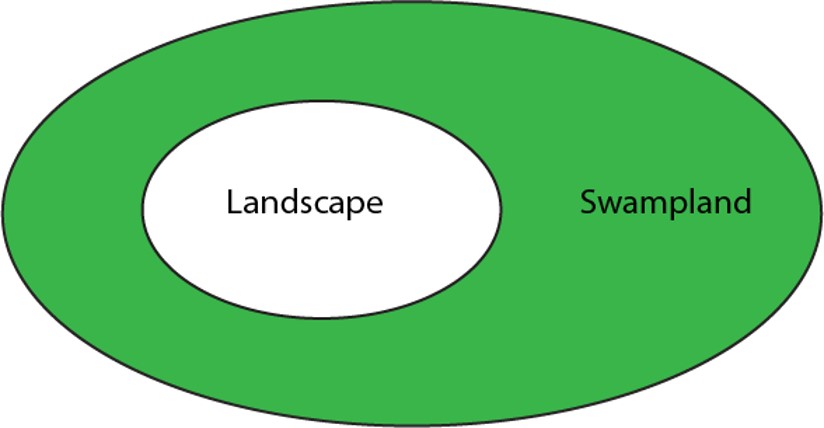}
\caption{The space of EFTs divided into the Landscape and the Swampland}
\label{fig}
\end{figure}

	The ultimate goal of the swampland program is to determine a full set of criteria that specify whether a given EFT lies in the landscape or the swampland. 
 At the moment, this goal is partially realized by the the so-called swampland conjectures. Some of these conjectures are old ideas, such as the statement that quantum gravity does not have global symmetries~\cite{Banks:1988yz,Banks:2010zn}. Recent work has resulted in a slew of new conjectures, along with a number of refinements and extensions. See~\cite{Brennan:2017rbf,Palti:2019pca,Yamazaki:2019ahj} for reviews on many of the conjectures and their applications.

We will focus on two of the conjectures in this work - the Swampland Distance~\cite{Ooguri:2006in,Klaewer:2016kiy,Baume:2016psm} and de Sitter~\cite{Obied:2018sgi} conjectures:
\begin{itemize}
\item The Swampland Distance Conjecture (SDC): As a field in an EFT traverses a distance $D$ in field space, the EFT breaks down as a tower of light modes appears with a mass scale

\begin{equation}
		m \sim \exp(-\beta D)
\label{eq:SDC}
\end{equation}
where $\beta$ an $\mathcal{O}(1)$ constant.
\item The de Sitter Conjecture (dSC): The scalar potential of an EFT in the landscape must satisfy
\begin{equation}
		|\nabla V| > cV
		\label{eq:dSC}
\end{equation}
						where $|\nabla V| =(g^{ij}\partial_i V\partial_j V)^{1/2} $, with $g_{ij}$ the metric in field space, 								and $c$ is an $\mathcal{O}(1)$ constant.
\end{itemize}
The most obvious implication of the dSC is that de Sitter vacua are excluded. Thus the observed dark energy\cite{Perlmutter:1998np,Riess:1998cb} should be caused by the rolling of a quintessence field rather than a cosmological constant~\cite{Obied:2018sgi,Agrawal:2018own}. Realistic quintessence models appear to be difficult to realize in string constructions~\cite{Cicoli:2018kdo}, but they can easily be embedded into supergravity theories~\cite{Brax:2009kd,Chiang:2018jdg}.\\

In this note, we show that the above conjectures can be used to establish a relation between inflation and quintessence models and how this relation affects the outlook for experiments measuring the dark energy equation of state. We will then repeat this analysis for single-field and multi-field inflationary models using a refinement of the dSC. 

\section{Cosmology and the Swampland}
	Let us consider single-field slow-roll inflation with the usual potential slow-roll parameters $\epsilon_V$ and $\eta_V$. If we make the approximation that $\epsilon_V$ is constant during inflation, then the field displacement of the inflaton can be simply related to the swampland parameter $c$ and the number of e-folds, $N_{e}$, via

\begin{equation}
	D = \int \sqrt{2\epsilon_V}dN_e \sim \sqrt{2\epsilon_V}N_e > cN_e
	\label{eq:sinf1}
\end{equation}

The distance conjecture places an upper bound on the magnitude of $D$: $D \le \mathcal{O}(1) = \alpha$. If we take $\alpha = 1$ and $N_e =50$, then $c < 0.02$. This indicates that the requirements of single-field inflationary physics are in tension with the notion that the swampland parameter $c$ is $\mathcal{O}(1)$. However, there is another reason why this result is troubling. If we consider the dark energy equation of state, the dSC can be used to obtain a lower bound on the deviation of w from -1:
\begin{equation}
					1+w = \frac{2(V_Q^\prime)^2}{(V_Q^\prime)^2+6V_Q^2} > \frac{2c^2}{6+c^2}\equiv \Delta(c).
					\label{eq:quint1}
\end{equation}
Where $V_Q$ is the potential of the quintessence field. For a given EFT, there exists a single value of the swampland parameter $c$ that appears in the dSC. Thus we can use our bound on $c$ from inflationary physics to see that

\begin{equation}
				\Delta(c) < \Delta(0.02) \sim 10^{-4}
				\label{eq:quint2}
\end{equation}

Current and future experiments~\cite{DES,HSC,DESI,PFS,LSST,Euclid,WFIRST} probe $\Delta$ to the percent level. It is unlikely that we will be able to probe $\Delta < 10^{-3}$ in the near future~\cite{Heisenberg:2018yae,Heisenberg:2018rdu,Akrami:2018ylq}. Thus it is possible that the dSC is correct but we would be unable to distinguish quintessence from a cosmological constant. 
  
We note that the above argument does not assume that inflation and quintessence arise from the dynamics of a single field. Rather, the fact that $c$ is universal in a given EFT means that the potential of the inflaton can be used to bound the quintessence equation of state.

\section{Cosmology and the Refined Swampland}
			The above argument hinges on the precise form of the dSC. From the low energy perspective, the dSC would seem to forbid the Higgs potential and perhaps even the QCD axion and QCD phase transition~\cite{Denef:2018etk, Murayama:2018lie,Choi:2018rze}. Refinements of the dSC were soon proposed that bypass these issues~\cite{Garg:2018reu,Ooguri:2018wrx}. We will take the Refined de Sitter Conjecture (RdSC) to have the following form:

\begin{itemize}
\item The Refined de Sitter Conjecture (RdSC): The scalar potential of an EFT in the landscape must satisfy
\begin{equation}
|\nabla V | > cV
\label{eq:rdsc1}
\end{equation}
\centering \textbf{or}
\begin{equation}
min(\nabla_i\nabla_j V) < -c^\prime V
\label{eq:rdsc2}
\end{equation}
\end{itemize}

Where the left hand side of Eq.~\ref{eq:rdsc2} is to be interpreted as taking the minimum eigenvalue of the Hessian matrix in an orthonormal frame. $c^\prime$ is a new parameter that is conjectured to also be $\mathcal{O}(1)$. This refined criterion greatly modifies the arguments from the previous section. In particular, the constraint on $c$ from inflation is relaxed since the RdSC applies a bound on the field displacement only for the number of e-folds that take place when the potential is convex~\cite{Fukuda:2018haz}:  $D  > cN_{convex}$. If $N_{convex}$ is less than the total number of e-folds, $N_{tot}$, then the bound from the previous section can be weakened. Consequences of the RdSC for quintessence models were also considered in~\cite{Agrawal:2018rcg}.

\subsection{Single-Field Inflation}
The discussion above suggests a natural strategy: construct inflaton potentials that satisfy the first part of the RdSC, Eq.~\ref{eq:rdsc1}, for $N_1$ e-folds and satisfy Eq.~(\ref{eq:rdsc2}) for the remaining $N_2 = N_{tot}- N_1$ e-folds. Let $(\epsilon_V^{(1)}, \eta^{(1)}_V)$ and $(\epsilon^{(2)}_V,\eta^{(2)}_V)$ be the slow roll parameters during the first and second parts of inflation, respectively. Then the SDC and RdSC impose the following constraints:\\

\noindent During the first $N_1$ e-folds of inflation, we must satisfy Eq.~\ref{eq:rdsc1}, or
\begin{equation}
\sqrt{2\epsilon^{(1)}_V} > c.
\end{equation}
The spectral tilt $n_s$ and Eq.~\ref{eq:rdsc2} provide a bound for $c^\prime$ during the second $N_2$ e-folds of inflation,  

\begin{equation}
	c^\prime < \frac{1}{2}\bigg(1-n_s(k)-6\epsilon^{(2)}_V\bigg).
	\label{eq:n2b}
\end{equation}

\noindent The SDC enforces

\begin{equation}
		\sqrt{2\epsilon_V^{(1)}}N_1 + \sqrt{2\epsilon_V^{(2)}} N_2 \le \alpha
\end{equation}
Finally, we must include the experimental upper bound on the tensor-to-scalar ratio $r$~\cite{Akrami:2018odb}: 
\begin{equation}
r<0.064. 
\label{eq:rb}
\end{equation}
For simplicity, we will assume that $\epsilon_V^{(2)}$ is negligible and drop it from the above constraints.
Note that we are incorporating the running of the spectral tilt in Eq.~\ref{eq:n2b}. Using the Planck analysis~\cite{Akrami:2018odb} and adding the errors in quadrature, we find

\begin{equation}
	n_s(k) = 0.9659 - 0.0041 \ln (k/k_*)\pm \sqrt{0.004^2 + (0.0067\ln(k/k_*))^2}
\end{equation}
with $k_*$ as the pivot scale. We take the 1$\sigma$ allowed lower end and so choose the negative sign. This results in the parameter space shown below.

\begin{figure}[!ht]
\hspace*{1cm}
\centerline{\includegraphics[width=0.7\linewidth]{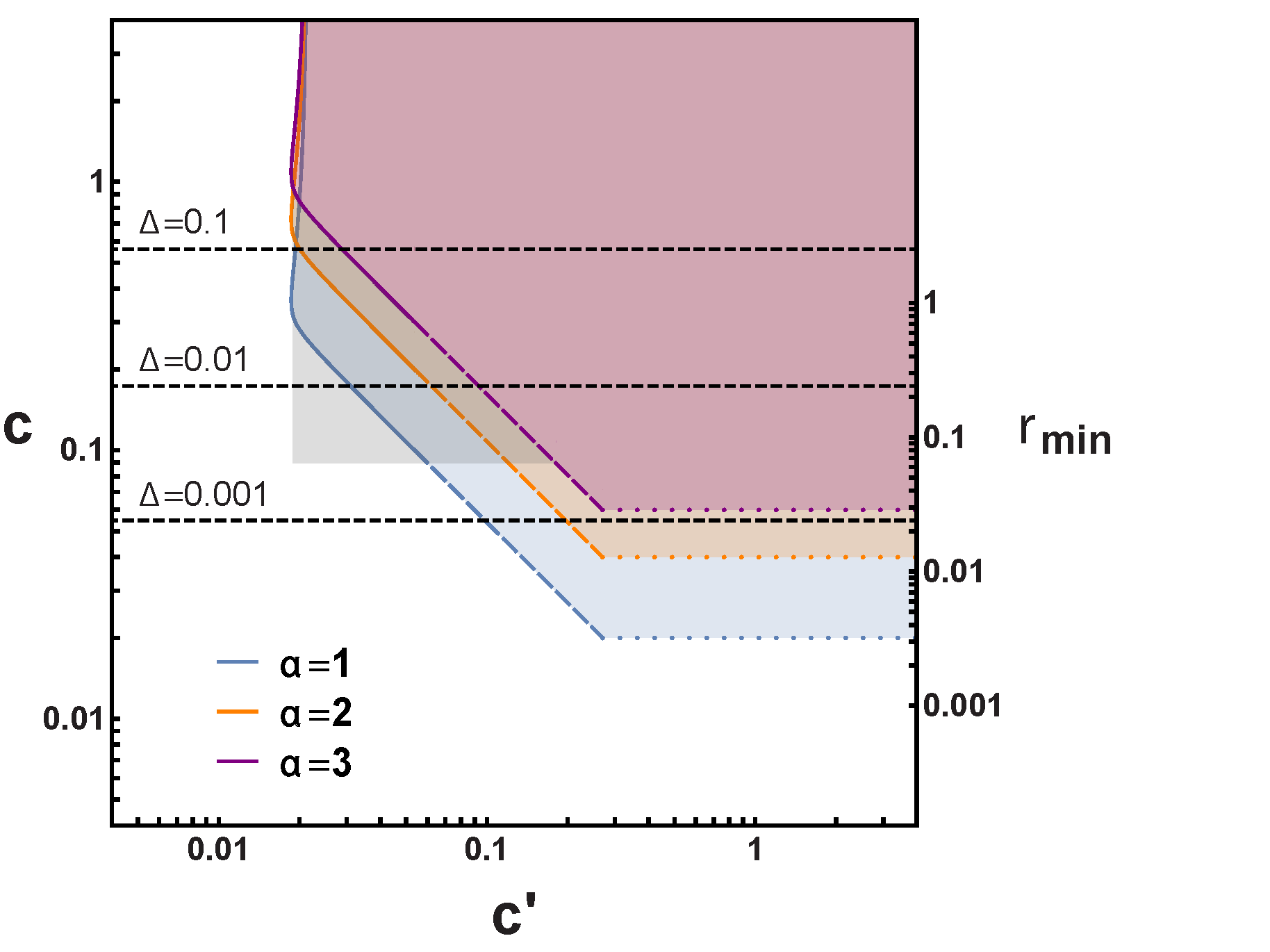}}
\caption[]{Bounds on $c$ and $c^\prime$ from single-field slow-roll inflation with $N_{tot} = 50$ and several values of $\alpha$.}
\label{fig:single}
\end{figure}

\vspace{5cm}

The three excluded parameter regions in Fig.~\ref{fig:single} correspond to relaxing the SDC constraint from $\alpha =1$ to $\alpha =3$ in integer steps. The data on the running of $n_s$ only goes up to $N_1 \sim10$, which is indicated by the solid line portions of the borders. The dashed border segments signify our extrapolation of the data on the running from $N_1\sim 10$ to $N_1\sim 50$. Finally, the dotted lines correspond to $N_1 = 50$ and represent the strict bound on $c$ from the dSC. The grey region is the upper bound on $r$, which only extends to $k > 0.002$ Mpc$^{-1}$~\cite{Akrami:2018odb}.

If the inflaton potential is concave down for almost the entire period of inflation, then quintessence can be forced to be observable since $c$ is essentially unconstrained by inflationary physics. Instead, if the inflaton potential satisfies Eq.~\ref{eq:rdsc1} for more than a handful of e-folds, then the potential observability of quintessence is dimished as the upper bound on $c$ requires $\Delta \lesssim 10^{-2}$. In this way, the  RdSC vastly improves experimental prospects of measuring quintessence so long as the inflaton potential is predominantly concave.

	\subsection{Multi-Field Inflation}
		  The role of multi-field scenarios in alleviating tension between inflation and the original de Sitter conjecture was  discussed in~\cite{Achucarro:2018vey}. Here we will examine how multi-field models can alter the above strict bounds on $c$ and $c^\prime$. In multi-field inflation, the Hubble slow roll parameters, ($\epsilon_H , \eta_H$), are not simply related to the potential slow roll parameters, ($\epsilon_V, \eta_V$). In particular, a two-field scenario has~\cite{Hetz:2016ics}

\begin{equation}
			\epsilon_V = \epsilon_H \left(1 + \frac{\Omega^2}{9H^2}\right)\\
\end{equation}
\begin{equation}
		12\eta_V = (c_s^{-2}-1)\frac{M^2}{H^2}+2\frac{M^2}{H^2}+3(4\epsilon_H-\eta_H) - 2\left(\left(\frac{M^2}{H^2}- \frac{3}{2}(4\epsilon_H-\eta_H)\right)^2+9(c_s^{-2}-1)\frac{M^2}{H^2}\right)^{1/2}
\end{equation}
where $\Omega$ is the local angular velocity describing the bend of the inflaton trajectory in the potential, $M$ is the mass of the direction orthogonal to the inflaton, and  $c_s = \left(1+\frac{4\Omega^2}{M^2}\right)^{-2}$ is the effective sound speed of fluctuations. We also have modifications to the spectral tilt and tensor-to-scalar ratio:

\begin{eqnarray}
					n_s &=& 1-2\epsilon_H - \eta_H - \frac{\dot{c}_s}{Hc_s}\\
					r &=& 16\epsilon_H c_s.
					\label{eq:r2}
\end{eqnarray}
 Since the RdSC constrains the potential slow-roll parameters, there is some room to try and accomodate slow-roll inflation, quintessence, and larger values of the swampland parameters $c$ and $c^\prime$. We will consider constant sound speeds so that $\dot{c}_s=0$. We will also set the scale of the heavy direction to be $M\sim H$. With these choices, the analysis is similar to the single-field case except that we have a new parameter in the form of $c_s$. The only subtlety is in the application of the SDC. We will take the conservative approach and assume the SDC applies to the length of the path that the inflaton follows and not just the geodesic distance between the starting and ending points. 

\begin{center}
\begin{figure}[!ht]
\hspace*{0.5cm}
\centerline{\includegraphics[width=0.7\linewidth]{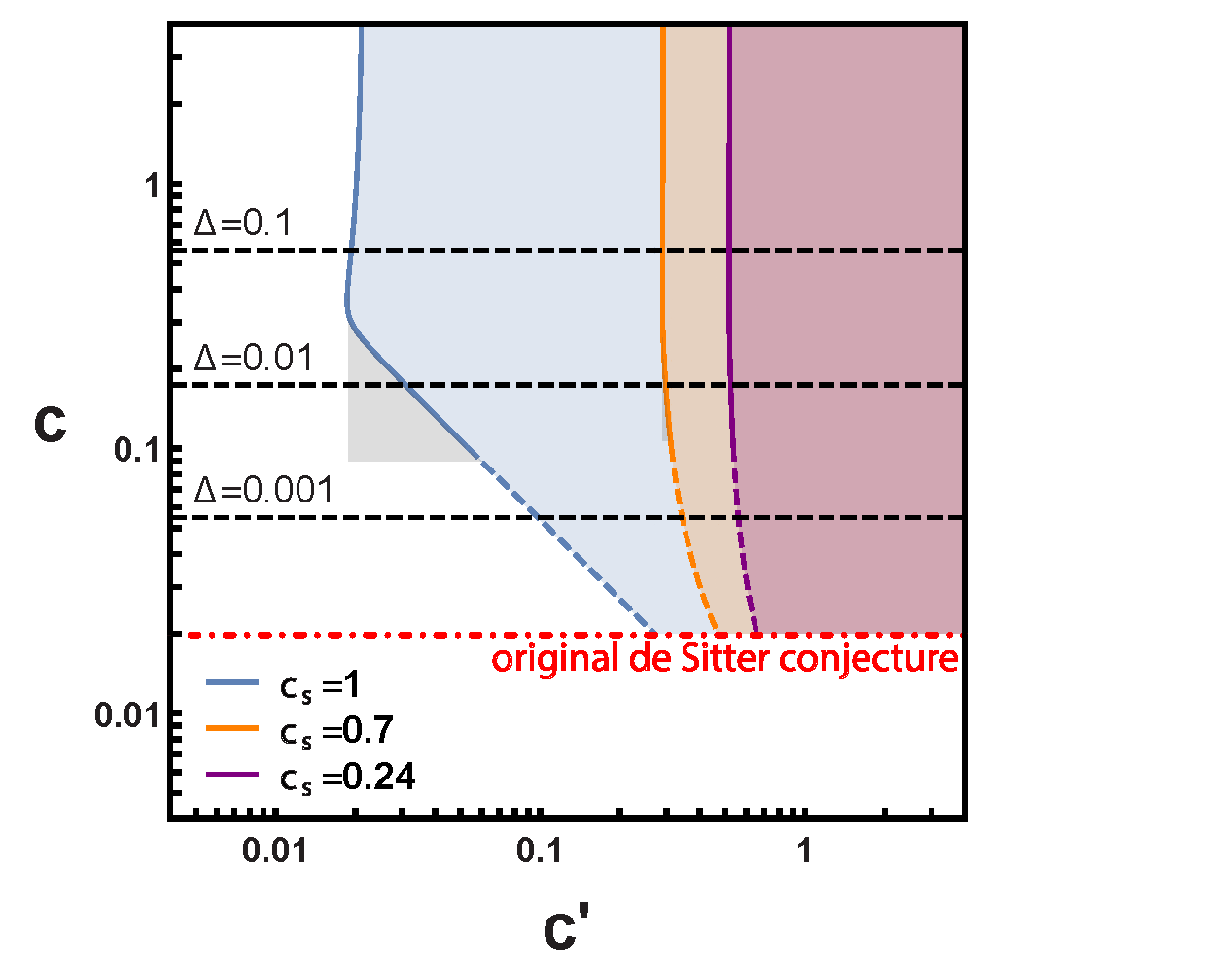}}
\caption[]{Bounds on $c$ and $c^\prime$ from multi-field slow-roll inflation with $N_{tot}= 50$ and  several values of $c_s$.}
\label{fig:double}
\end{figure}
\end{center}

Many aspects of Fig.~\ref{fig:double} are similar to those in Fig.~\ref{fig:single}. The borders of the excluded regions are again broken into solid, dashed, and dotted lines corresponding to $N_1 \lesssim 10$, $10\lesssim N_1\lesssim 50$, and $N_1=50$, respectively. However, here we have fixed $\alpha = 1$ and the three regions correspond to lowering the sound speed from unity. 
For $c_s = 0.7,0.24$ the grey regions from the bound on $r$ are altered due to the modified expression for $r$ in  Eq.~\ref{eq:r2}.
	It is clear that lowering the sound speed allows for much larger values of $c$ and $c^\prime$ compared to the single-field case. Furthermore, the inflaton potential can be convex for $\mathcal{O}(10)$ e-folds and still allow for observable quintessence.  
Note that we are working with $c_s$ values far from the lower bound of $c_s \ge 0.024$~\cite{Ade:2015rim}. 

\clearpage

\section{Conclusion}
	In this note we have considered how inflationary models affect the observability of quintessence due to the swampland conjectures. The original de Sitter conjecture, coupled with inflation, allows for a situation in which dark energy could truly be caused by a quintessence field but we would be unable to conclusively prove this via experimental searches. We have shown that this situation is improved in light of the refined de Sitter conjecture. In a single-field inflationary scenario, the freedom to satisfy the second part of the RdSC allows for the reconciliation of inflationary physics with quintessence models that must be observable. However, there is significant tension between inflation and the RdSC since $\mathcal{O}(0.01)$ values of $c^\prime$ are required. On the other hand, in multi-field inflationary scenarios, inflation can coexist with larger $c$ and $c^\prime$ values so long as the sound speed is reduced from unity. The utility of multi-field scenarios as a means to address both inflation and the  swampland was also considered in~\cite{Achucarro:2018vey,Damian:2018tlf,Aragam:2019khr,Bravo:2019xdo}. The above results imply that the experimental efforts to better understand dark energy are not only worthwhile in their own right, but they could also provide glimpses into quantum gravity through the window furnished by the swampland program. 

\section{Acknowledgements}
	The author would like to thank the ICISE for providing hospitality and a stimulating environment during Rencontres du Vietnam 2019. The author is partially supported by U.S. DOE Contract DE-AC02-05CH11231.
\section{References}

\bibliography{Refs_2}

\end{document}